\documentclass[twocolumn,showpacs,amsmath,amssymb]{revtex4}

\usepackage{graphicx}
\usepackage{dcolumn}
\usepackage{bm}

\renewcommand{\vec}[1]{{\mathbf #1}}

\begin{document}

\title{Comment on ``Scaling behavior of classical wave transport in mesoscopic media at the localization transition''}

\author{N. Cherroret}
\author{S.E. Skipetrov}
\author{B.A. van Tiggelen}
\affiliation{
Universit\'{e} Joseph Fourier, Laboratoire de Physique et Mod\'{e}lisation des Milieux Condens\'{e}s,\\
CNRS UMR 5493, B.P. 166, 25 rue des Martyrs, Maison des Magist\`{e}res,
38042 Grenoble Cedex 09, France}

\date{\today}

\begin{abstract}
We emphasize the importance of the position dependence of the diffusion coefficient $D(\vec{r})$ in the self-consistent theory of localization and argue that the scaling law $T \propto \ln L/L^2$ obtained by
Cheung and Zhang [Phys. Rev. B \textbf{72}, 235102 (2005)] for the average transmission coefficient $T$ of a disordered slab of thickness $L$ at the localization transition is an artifact of replacing $D(\vec{r})$ by its harmonic mean. The correct scaling $T \propto 1/L^2$ is obtained by properly treating the position dependence of $D(\vec{r})$.
\end{abstract}

\maketitle

In a recent paper \cite{cheung05} Cheung and Zhang (CZ) apply the self-consistent (SC) theory of localization to study the transmission of waves through a slab of disordered medium at the Anderson localization transition. The SC theory is a powerful tool to deal with the phenomenon of Anderson localization, but its application to disordered media of finite size requires some care. In the original papers by Vollhardt and W\"{o}lfle  \cite{voll80}, the size $L$ of disordered sample was acknowledged using a lower cut-off in the integration over momentum. Despite the obvious crudeness of this approach, it was sufficient to recover the main results of the scaling theory of localization \cite{abrahams79} and added a great physical insight into the phenomenon of disorder-induced localization. Later on, Van Tiggelen {\em et al.} \cite{tiggelen00} argued that in a medium of finite size the SC theory naturally leads to a position dependence of the diffusion coefficient $D(\vec{r})$. This adapted SC theory was successfully applied to study coherent backscattering \cite{tiggelen00} and dynamics \cite{skip04,skip06} of localized waves. Microscopic justifications for position dependence of $D$ have been recently presented based on the diagrammatic \cite{cherroret08} and field-theoretic \cite{tian08} calculations.

CZ propose a way of overcoming technical difficulties caused by the position dependence of $D(\vec{r})$ \cite{cheung05} (see also \cite{cheung04}). They average the equation for $1/D(\vec{r})$, their Eq.\ (1), over the sample volume, thus replacing $D(\vec{r})$ by its harmonic mean ${\bar D}$. In this Comment we argue that although such an approach can be justified in the weak localization regime \cite{cheung04}, it is not adequate at the mobility edge and in the Anderson localization regime. In particular, our calculations that properly treat the position dependence of  $D(\vec{r})$, do not confirm the scaling law $T \propto \ln L/L^2$ found by CZ for the transmission coefficient $T$ of a disordered slab of thickness $L$ at the mobility edge. Instead, we find $T \propto 1/L^2$ in agreement with the  scaling theory of localization \cite{abrahams79}.

\begin{figure}
\includegraphics[width=\columnwidth]{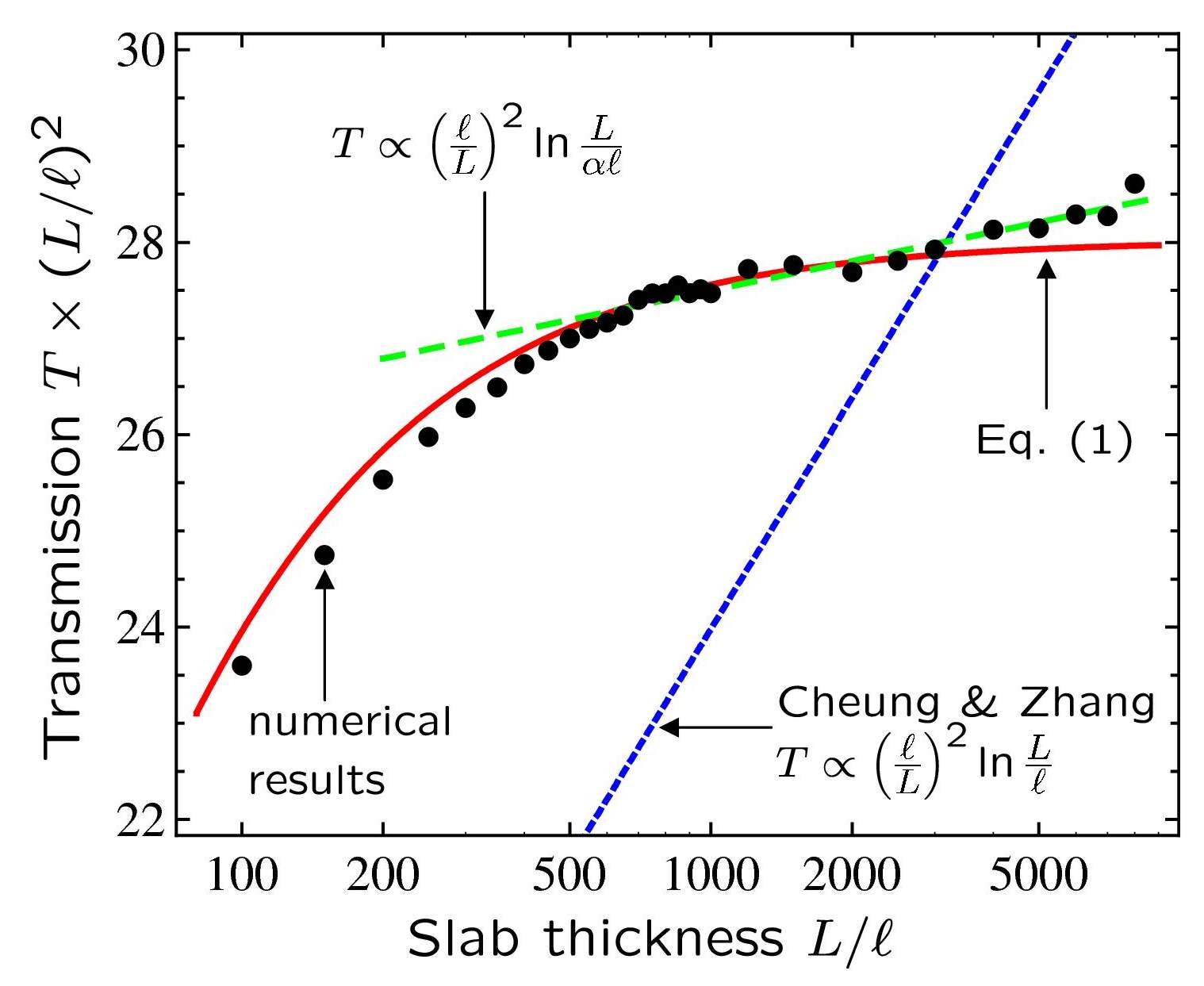}
\caption{\label{fig1} Average transmission coefficient $T$ of a disordered slab of thickness $L$ at the Anderson localization transition. Circles were obtained from the self-consistent theory of localization with a position-dependent diffusion coefficient $D(z)$ by numerical solution \cite{skip06}. The solid red line is a fit to the numerical results using Eq.\ (\ref{t}) with $D(0)/D_B = 0.82$ and $z_c = 4.2 \ell$. The dotted blue and dashed green lines are fits to numerical data for $L/\ell > 10^3$ using $T \propto (\ell/L)^2 \ln(L/\ell)$ and $T \propto (\ell/L)^2 \ln(L/\alpha \ell)$, respectively. We obtain $\alpha \simeq 7.37 \times 10^{-25}$ in the latter case.
}
\end{figure}

To study the scaling of the average transmission coefficient $T$ with the thickness $L$ of disordered slab, we solve the two equations of SC theory --- Eqs.\ (1) and (2) of Ref.\ \cite{skip06} with $\Omega = 0$ (stationary regime) and $k \ell = 1$ (mobility edge) \footnote{In contrast to Ref.\ \cite{cheung05}, we use a 2D cutoff $q_{\mathrm{max}} = \mu/\ell$ in the integration over momentum $\vec{q}$ in Eq.\ (2) of Ref.\ \cite{skip06} and adjust $\mu = 1/3$ to obtain the mobility edge in the infinite medium at $k \ell = 1$. The need for a cutoff arises from the failure of the small-$q$ approximation implied by Eq.\ (1) of Ref.\ \cite{skip06} for $q > 1/\ell$ (small distances). The exact way of applying the cutoff (2D or 3D cutoff, exact value of $\mu$, etc.) does not affect scaling with (large) $L$ as far as the cutoff is consistent with the definition of the mobility edge.} --- numerically. We use the same boundary conditions and the same method of numerical solution as in Ref.\ \cite{skip06} and vary the thickness of the slab $L$ from  $10^2 \ell$ to $8 \times 10^3 \ell$. Here $k$ is the wave number of the wave and $\ell$ is the mean free path due to disorder. Our results are presented in Fig.\ \ref{fig1} by circles. The red solid line in Fig.\ \ref{fig1} shows
\begin{eqnarray}
T = \left( \frac{\ell}{L} \right)^2
\frac{2 + 4 \frac{z_c}{\ell} \left[ 1 + \frac{D(0)}{D_B} \frac{z_0}{\ell} \right]
}{1 + 4 \frac{z_c}{L} \left[ 1 + 2 \frac{D(0)}{D_B} \frac{z_0}{L} \right]}
\label{t}
\end{eqnarray}
that we obtained by assuming $D(z) = D(0)/(1 + {\tilde z}/z_c)$ with ${\tilde z} = \min(z, L-z)$ as suggested by Van Tiggelen \emph{et al.} \cite{tiggelen00}. Here $D_B$ is the diffusion coefficient in the absence of macroscopic interferences (i.e. in the limit of $k \ell \gg 1$). $D(0)/D_B$ was determined directly from the numerical results at a sufficiently large $L = 10^3 \ell$, whereas $z_c$ was a free fit parameter. We used $z_0 = \frac{2}{3} \ell$, corresponding to no internal reflections at the sample boundaries. Deviations of the fit from the numerical results do not exceed 3\% in the whole range of considered $L$'s, which supports the validity of Eq.\ (\ref{t}) and its underlying model for $D(z)$. The inaccuracy of the latter model in the middle of the slab cause deviations at small $L < 10^3 \ell$, whereas deviations at large $L > 4 \times 10^3 \ell$ are mostly due to the extremely slow convergence of our computational algorithm for thick slabs and would, most likely, disappear if more computer time were available. We note that $T \times (L/\ell)^2$ grows with $\ln(L/\ell)$ for $L < 10^3 \ell$, but then saturates at a constant level for larger $L$, suggesting $T \propto (\ell/L)^2$ in the limit of large $L$.

Neither the ensemble of numerical results of Fig.\ \ref{fig1}, nor its small- or large-$L$ parts can be fit by $T = \mathrm{const} \times (\ell/L)^2 \ln(L/\ell)$ proposed by CZ. This is easy to see from Fig.\ \ref{fig1} where we show a fit of the above equation to our numerical data for $L/\ell > 10^3$ (dotted blue straight line). It is clear that the fast growth of $T \times (L/\ell)^2$ with $\ln(L/\ell)$ predicted by CZ is not supported by our numerical calculations: the numerical results only show an increase of 20\% in the range of $L/\ell = 100$--8000 and 4\% in the range $L/\ell = 1000$--8000, whereas the result of CZ increases by 100\% and 30\%, respectively. For large $L > 10^3 \ell$, a reasonable fit can be achieved by $T \propto (\ell/L)^2 \ln (L/\alpha \ell)$. The result of CZ would correspond to $\alpha \sim 1$, whereas a fit to the numerical data yields $\alpha \sim 10^{-24} \ll 1$. This value is unphysically small and implies existence of length scales that are 24 orders of magnitude shorter than the mean free path $\ell$. We therefore conclude that our numerical results exclude the possibility of logarithmic scaling of $T \times (L/\ell)^2$ with $L/\ell$. Appearance of this scaling in Ref.\ \cite{cheung05} should then be an artifact of replacing $D(\vec{r})$ by its harmonic mean.

In conclusion, we have shown the importance of properly treating the position dependence of the diffusion coefficient $D(\vec{r})$ in the SC theory of localization. In particular, replacing $D(\vec{r})$ by its harmonic mean leads to an incorrect scaling law for the transmission coefficient $T$ with the thickness $L$ of disordered slab at the mobility edge.  The correct scaling law $T \propto 1/L^2$ is obtained by solving SC equations with a position dependent $D(\vec{r})$.

The computations presented in this paper were performed on the cluster HealthPhy (CIMENT, Grenoble).
S.E.S. acknowledges financial support from the French ANR (project No. 06-BLAN-0096 CAROL) and the French Ministry of Education and Research.



\end{document}